\documentclass[prl,twocolumn,aps,superscriptaddress,preprintnumbers,letterpaper,showpacs]{revtex4}
\usepackage{amsmath,amssymb}
\usepackage{epsfig}
\usepackage{graphicx}
\usepackage{amsmath}
\usepackage{amsfonts}
\usepackage{epstopdf}

\begin{document}
\renewcommand{\vec}{\boldsymbol}
\newcommand{\mc}{M_{\mathrm{c}}}
\newcommand{\mn}{M_{\mathrm{n}}}
\newcommand{\mnc}{M_{\mathrm{nc}}} 
\newcommand{\mr}{M_{\mathrm{R}}}
\newcommand{\gma}{\gamma}
\newcommand{\gmat}{\tilde{\gamma}}
\newcommand{\pc}{\vec{p}_{c}}
\newcommand{\pn}{\vec{p}_{n}}
\newcommand{\bes}{{}^{7}\mathrm{Be}}
\newcommand{\besst}{{}^{7}\mathrm{Be}^{\ast}}
\newcommand{\be}{{}^{8}\mathrm{B}}
\renewcommand{\S}[2]{{}^{#1}S_{#2}}
\renewcommand{\P}[2]{{}^{#1}P_{#2}}
\newcommand{\gone}{g_{(\S{3}{1})}}
\newcommand{\gtwo}{g_{(\S{5}{2})}}
\newcommand{\gthree}{g_{(\S{3}{1}^{*})}}
\newcommand{\aone}{a_{(\S{3}{1})}}
\newcommand{\atwo}{\ensuremath{a_{(\S{5}{2})}}}
\newcommand{\hone}{h_{(\P{3}{2})}}
\newcommand{\htwo}{h_{(\P{5}{2})}}
\newcommand{\hpt}{h_{t}}
\newcommand{\hthree}{h_{(\P{3}{2}^{*})}}
\newcommand{\honet}{\tilde{h}_{(\P{3}{1})}}
\newcommand{\htwot}{\tilde{h}_{(\P{5}{1})}}
\newcommand{\Xone}{{X}_{(\S{3}{1})}}
\newcommand{\Xtwo}{{X}_{(\S{5}{2})}}
\newcommand{\Xonet}{\tilde{X}_{(\S{3}{1})}}
\newcommand{\Xtwot}{\tilde{X}_{(\S{5}{2})}}
\newcommand{\V}[1]{\vec{V}_{#1}}
\newcommand{\fdu}[2]{{#1}^{\dagger #2}}
\newcommand{\fdd}[2]{{#1}^{\dagger}_{#2}}
\newcommand{\fu}[2]{{#1}^{#2}}
\newcommand{\fd}[2]{{#1}_{#2}}
\newcommand{\T}[2]{T_{#1}^{\, #2}}
\newcommand{\e}{\vec{\epsilon}}
\newcommand{\es}{\e^{*}}
\newcommand{\cw}[2]{\chi^{(#2)}_{#1}}
\newcommand{\cwc}[2]{\chi^{(#2)*}_{#1}}
\newcommand{\cwf}[1]{F_{#1}}
\newcommand{\cwg}[1]{G_{#1}}

\newcommand{\ke}{k_{E}}

\newcommand{\kest}{k_{E\ast}}

\newcommand{\kc}{k_{C}}
\newcommand{\upartial}[1]{\partial^{#1}}
\newcommand{\dpartial}[1]{\partial_{#1}}
\newcommand{\etae}{\eta_{E}}
\newcommand{\etab}{\eta_{B}}
\newcommand{\etaest}{\eta_{E\ast}}
\newcommand{\etabst}{\eta_{B\ast}}
\newcommand{\vecpt}[1]{\hat{\vec{#1}}}
\newcommand{\uY}[2]{Y_{#1}^{#2}}
\newcommand{\dY}[2]{Y_{#1 #2}}

\def\lsim{\mathrel{\rlap{\lower4pt\hbox{\hskip1pt$\sim$}}
    \raise1pt\hbox{$<$}}}         
\def\gsim{\mathrel {\rlap{\lower4pt\hbox{\hskip1pt$\sim$}}
    \raise1pt\hbox{$>$}}}         

\title{Halo effective field theory constrains the solar ${}^7{\rm Be} + p \rightarrow {}^8{\rm B} + \gamma$ rate}


%
\author{Xilin Zhang} \email{xilinz@uw.edu}
\affiliation{Physics Department, University of Washington, 
Seattle, WA 98195, USA} 
\affiliation{Institute of Nuclear and Particle Physics and Department of
Physics and Astronomy, Ohio University, Athens, OH\ \ 45701, USA}
\author{Kenneth M.~Nollett} \email{nollett@mailbox.sc.edu}
\affiliation{Department of Physics and Astronomy, 
University of South Carolina,
712 Main Street, Columbia, South Carolina 29208, USA} 
\affiliation{Department of Physics, San Diego State University,
5500 Campanile Drive, San Diego, California 92182-1233, USA} 
\affiliation{Institute of Nuclear and Particle Physics and Department of
Physics and Astronomy, Ohio University, Athens, OH\ \ 45701, USA}

\author{D.~R.~Phillips} \email{phillid1@ohio.edu}
\affiliation{Institute of Nuclear and Particle Physics and Department of
Physics and Astronomy, Ohio University, Athens, OH\ \ 45701, USA}

\date{July 2015}

\begin{abstract}
We report an improved low-energy extrapolation of the cross section
for the process $^7\mathrm{Be}(p,\gamma)^8\mathrm{B}$, which
determines the $^8$B neutrino flux from the Sun.  Our extrapolant is
derived from Halo Effective Field Theory (EFT) at next-to-leading
order.  We apply Bayesian methods to determine the EFT parameters and
the low-energy $S$-factor, using measured cross sections and
scattering lengths as inputs. Asymptotic normalization coefficients of
$^8$B are tightly constrained by existing radiative capture data, and
contributions to the cross section beyond external direct capture are
detected in the data at $E < 0.5$ MeV.  Most importantly, the
$S$-factor at zero energy is constrained to be $S(0)= 21.3\pm 0.7$
\mbox{eV b}, which is an uncertainty smaller by a factor of two than
previously recommended. That recommendation was based on the full range for $S(0)$ obtained among a discrete set of models judged to  be reasonable. 
In contrast, Halo EFT subsumes all models into a controlled low-energy approximant, where they are characterized by nine parameters at next-to-leading order. These are fit to data, and marginalized over via Monte Carlo integration to produce the improved prediction for $S(E)$.


\end{abstract}

\pacs{25.20.-x, 25.40.Lw, 11.10.Ef, 21.10.Jx, 21.60.De}

\maketitle

{\em Introduction---}
A persistent challenge in modeling the Sun and other stars is the need
for nuclear cross sections at very low
energies~\cite{Adelberger:2010qa,rolfs-rodney}.  Recent years have
seen a few measurements at or near the crucial ``Gamow peak'' energy
range for the Sun~\cite{luna,Adelberger:2010qa}, but cross sections at
these energies are so small that data almost always lie at higher
energies, where experimental count rates are larger.  The bulk of the
data must be extrapolated to the energies of stellar interiors using
nuclear reaction models.

The models available for extrapolation also have limitations.
Qualitatively correct models of nonresonant radiative capture
reactions, with reacting nuclei treated as interacting particles, have
been available since the mid-1960s~\cite{christyduck}.  However, these
models suffer from weak input constraints and dependence on \textit{ad
  hoc} assumptions like the shapes of potentials.  Developing models
with realistically interacting nucleons as their fundamental degrees
of freedom is currently a priority for the theoretical community, but
progress is slow, and models remain
incomplete~\cite{Descouvemont:2004hh,Navratil:2011sa}.  {\it Ab initio} 
calculations employing modern nuclear forces may yield tight constraints
in the future.

For the $^7\mathrm{Be}(p,\gamma)^8\mathrm{B}$ reaction -- which
determines the detected flux of $^8$B decay neutrinos from the Sun --
the precision of the astrophysical $S$-factor at solar energies ($\sim
20$ keV) is limited by extrapolation from laboratory energies of
typically 0.1--0.5 MeV.  A recent evaluation \cite{Adelberger:2010qa}
found the low-energy limit $S(0) = 20.8 \pm 0.7\pm 1.4$ \mbox{eV b},
with the first error reflecting the uncertainties of the measurements.
The second accounts for uncertainties in extrapolating those data.  It
was chosen to cover the full variation among a few extrapolation
models thought to be plausible.  Since the differences among $S(E)$
shapes for different models were neither well-understood nor
represented by continuous parameters, no goodness-of-fit test was used
for model selection.

Halo EFT~\cite{vanKolck:1998bw,Kaplan:1998tg,Kaplan:1998we,Bertulani:2002sz,Bedaque:2003wa,Hammer:2011ye,Rupak:2011nk,Canham:2008jd,Higa:2008dn,Ryberg:2013iga}, provides a simple,
transparent, and systematic way to organize the reaction theory
needed for the low-energy extrapolation.  The $^7\mathrm{Be}+p$ system is modeled as two
interacting particles and described by a Lagrangian expanded in powers
of their relative momentum, which is small compared with other
momentum scales in the problem.  The point-Coulomb part of the
interaction can be treated exactly, and the form of the strong
interaction is fully determined by the order at which the Lagrangian
is truncated~\cite{Ryberg:2013iga,Zhang:2014zsa,Zhang:2015,Ryberg:2014exa}.  The
coupling constants of the Lagrangian are determined by matching to
experiment. This is similar in spirit and in many quantitative details to 
traditional potential model or $R$-matrix approaches. However, it avoids
some arbitrary choices (like Woods-Saxon shapes or matching radii) of 
these models, is organized explicitly as a low-momentum power
series, and allows quantitative estimates of the error arising from
model truncation.  

The low-energy $S$-factor for
$^7\mathrm{Be}(p,\gamma)^8\mathrm{B}$ consists entirely of electric-dipole 
($E1$) capture from $s$- and $d$-wave initial states to
$p$-wave final states (which dominate $^7\mathrm{Be}+p$ configurations
within $^8$B).  All models are dominated by ``external direct
capture,'' the part of the $E1$ matrix element arising in the
tails of the wave function (out to 100 fm and
beyond)~\cite{christyduck,jennings98}.  Models differ in how they
combine the tails of the final state with phase shift information and
in how they model the non-negligible contribution from short-range,
non-asymptotic regions of the wave functions. 

Halo EFT includes these mechanisms, and can describe $S(E)$ over the
low-energy region (LER) at $E < 0.5$ MeV.  Beyond 0.5 MeV,
higher-order terms could be important, and resonances unrelated to the
$S$-factor in the Gamow peak appear.  Compared with a potential model,
the EFT has about twice as many adjusted parameters, too many to
determine uniquely with existing data.  However, calculations of the
solar neutrino flux do not require that all parameters be known: it is
enough to determine $S(18\pm 6~\mathrm{keV})$. We fit the amplitudes
of recently computed next-to-leading-order (NLO)
terms~\cite{Zhang:2015} in $E1$ capture to the experimental $S(E)$
data in the LER. We then use Bayesian methods to propagate the (theory
and experimental) uncertainties and obtain a rather precise result for
$S(20~\mathrm{keV})$.

{\em EFT at NLO---}
The EFT amplitude for $E1$ capture is organized in an expansion in the ratio of low-momentum and high-momentum scales, $k/\Lambda$. $\Lambda$ is set by the $\bes$ binding energy relative to the ${}^3\mathrm{He}+{}^4\mathrm{He}$ threshold, $1.59$ MeV,  so $\Lambda \approx 70$ MeV, corresponding to
a co-ordinate space cutoff of $\approx 3$ fm. Physics at distances shorter than this is subsumed into contact operators in the Lagrangian. The $\be$ ground state, which is 0.1364(10) MeV below the $\bes$-p scattering continuum~ \cite{AME2012I, AME2012II}, is a shallow p-wave bound state in our EFT: it is bound by contact operators but the wave-function tail should be accurately represented. To ensure this we also include the $J^\pi=\frac{1}{2}^-$ bound excited state of $\bes$ in the theory. $\bes^*$ is 0.4291 MeV above the ground state; the configuration containing it and the proton is significant in the $\be$ ground state~\cite{Zhang:2014zsa}. The large ($\sim 10$ fm) $\bes$-p scattering lengths play a key role in the low-energy dynamics; s-wave rescattering in the incoming channels must be accurately described. This also
requires that the Coulomb potential be iterated to all orders when computing the scattering and bound state wave functions~\cite{Zhang:2014zsa,Ryberg:2014exa}. Indeed $Z_{\bes} Z_p \alpha_{em} M_{p} \approx k_C = 27$ MeV while the binding momentum of $\be$ is 15 MeV, so these low-momentum scales are well separated from $\Lambda$. We generically denote them by $k$, and anticipate that $k/\Lambda \approx 0.2$. Since the EFT  incorporates all dynamics at momentum scales $< \Lambda$ its radius of convergence is larger than other efforts at systematic expansions of this $S$-factor~\cite{WilliamsKoonin81,Baye:2000ig, Baye:2000gi,Baye:2005, Jennings:1998qm,Jennings:1998ky,Jennings:1999in, Cyburt:2004jp, Mukhamedzhanov:2002du}. 

The leading-order (LO) amplitude includes only external direct capture. As the $\bes$ ground state is $\frac{3}{2}^-$ there are two possible total spin channels, denoted here by $s=1,2$. They correspond, respectively, to 
 $\S{3}{1}$ and $\S{5}{2}$ components in the incoming scattering state, and $\P{3}{2}$ and $\P{5}{2}$ configurations in  $\be$.
The parameters that appear at LO are the two asymptotic normalization coefficients (ANCs), $C_{s}$, for the $\bes$-p configuration in $\be$ in each of the spin channels, together with the corresponding s-wave scattering lengths, $a_{s}$~\cite{Zhang:2013kja,Zhang:2014zsa,Ryberg:2014exa}.  The NLO result for $S(E)$, full details of which will be given elsewhere~\cite{Zhang:2015}, can be written as:
\begin{widetext}

\begin{eqnarray}
S(E)&=&f(E) \sum_{s} C_{s}^2 
\bigg[ \big\vert \mathcal{S}_\mathrm{EC} \left(E;\delta_s(E)\right) 
+ \overline{L}_{s} \mathcal{S}_\mathrm{SD} \left(E;\delta_s(E)\right)  
+ \epsilon_{s} \mathcal{S}_\mathrm{CX}\left(E;\delta_s(E)\right) \big\vert^2 
+|\mathcal{D}_\mathrm{EC}(E)|^2 \bigg] \ . 
\end{eqnarray}

Here, $f(E)$ is an overall normalization composed of final-state phase space over incoming flux ratio, dipole radiation coupling strength, and a factor related to Coulomb-barrier penetration~\cite{Zhang:2014zsa}.
$\mathcal{S}_\mathrm{EC}$ is proportional to the  spin-$s$ $E1$ \cite{Walkecka1995book, Zhang:2014zsa, Zhang:2013kja} external direct-capture matrix element between continuum $\bes$--p s-wave and $\be$ ground-state wave functions. $\mathcal{S}_\mathrm{CX}$ is the contribution from capture with core excitation, i.e.~into the $\bes^*$-p component of the ground state. Its strength is parameterized by $\epsilon_s$. Since $\bes^*$ is spin-half this component only occurs for $s=1$, so $\epsilon_2=0$. Because the inelasticity in  $\bes$-p s-wave scattering is small \cite{Navratil:2010jn, Navratil:2011sa} it is an NLO effect. 

Short-distance contributions, $\mathcal{S}_\mathrm{SD}$, are also
NLO. They originate from NLO contact terms in the EFT Lagrangian
\cite{Zhang:2015} and account for corrections to the LO result arising
from the $E1$ transition at distances $\lsim 3$ fm. The size of these
is set by the parameters $\overline{L}_s$, which must be fit to data.
$\mathcal{S}_\mathrm{EC}$, $\mathcal{S}_\mathrm{SD}$, and
$\mathcal{S}_\mathrm{CX}$ are each functions of energy, $E$, but
initial-state interactions mean they also depend on the s-wave phase
shifts $\delta_s$. At NLO we parametrize $\delta_s(E)$ by the
Coulomb-modified effective-range expansion up to second order in
$k^2$, i.e., we include the term proportional to $r_s k^2$, with $r_s$
the effective range (see supplemental
material)~\cite{Higa:2008dn,Koenig:2012bv,GoldbergerQM}. Finally,
$\mathcal{D}_\mathrm{EC}$ is the $E1$ matrix element between the d-wave scattering
state and the $\be$ bound-state wave function. It is not affected by
initial-state interactions up to NLO, and hence is the same for
$s=1,\,2$ channels and introduces no new parameters.
This leaves us with $9$ parameters in all: $C_{1,2}^2$, $a_{1,2}$ at LO and five more at NLO: $r_{1,2}$, $\overline{L}_{1,2}$, and $\epsilon_1$~\cite{Zhang:2015}.

\end{widetext}

{\em Data---} The 42 data points in our analysis come from all modern experiments with more than one data point for the direct-capture $S$-factor in the LER:
Junghans {\it et.al., } (experiments ``BE1'' and ``BE3'') \cite{Junghans:2010zz},  Filippone {\it et.al.,} \cite{Filippone:1984us}, Baby {\it et.al.,} \cite{Baby:2002hj, Baby:2002ju}, and Hammache {\it et.al.,} (two measurements published in 1998 and 2001) \cite{Hammache:1997rz, Hammache:2001tg}.
Ref.~\cite{Adelberger:2010qa} summarizes these experiments, and the common-mode errors (CMEs) we assign are given in the supplemental material. All data are for energies above $0.1$ MeV. We subtracted the $M1$ contribution of the $\be$ $1^{+}$ resonance from the data using the resonance parameters of Ref.~\cite{Filippone:1984us}. This has negligible impact for $E \leq 0.5$ MeV, so we retain only points in this region, thus eliminating the resonance's effects.

{\em Bayesian analysis---} To extrapolate $S(E)$ we must use these data to constrain the EFT parameters. We compute the posterior probability distribution function (PDF) of the parameter vector $\vec{g}$ given data, $D$, our theory, $T$, and prior information, $I$. To account for the common-mode errors 
in the data we introduce data-normalization corrections, $\xi_i$. We then employ Bayes' theorem to write the desired PDF as:
\begin{equation}
{\rm pr} \left(\vec{g},\{\xi_i\} \vert D;T; I \right)  
=
{\rm pr} \left(D \vert \vec{g},\{\xi_i\};T; I \right) {\rm pr} \left(\vec{g},\{\xi_i\} \vert I \right) , \label{eqn:bayesian1}
\end{equation}
with the first factor proportional to the likelihood:
$$
\ln {\rm pr} \left(D \vert \vec{g},\{\xi_i\};T;I \right)  =  c - \sum_{j=1}^N \frac{\left[ (1-\xi_j)S(\vec{g}; E_j)-D_j\right]^2}{2  \sigma_j^{2}},
$$
where $S(\vec{g};E_j)$ is the NLO EFT $S$-factor at the energy $E_j$ of the $j$th data point $D_j$, whose statistical uncertainty is $\sigma_j$. The constant $c$ ensures ${\rm pr} \left(\vec{g},\{\xi_i\} \vert D;T; I \right)$ is normalized. Since the CME affects all data from a particular experiment in a correlated way there are only five parameters $\xi_i$: one for each experiment. 

In Eq.~(\ref{eqn:bayesian1}) ${\rm pr}\left( \vec{g}, \{ \xi_i\},\vert
I \right)$ is the prior for $\vec{g}$ and $\{\xi_i\}$. We choose
independent Gaussian priors for each data set's $\xi_i$, all centered
at $0$ and with width equal to the assigned CMEs. We also choose
Gaussian priors for the s-wave scattering lengths $\left(a_{1},\,
a_2\right)$, with centers at the experimental values of
Ref.~\cite{Angulo2003}, $\left(25,\, -7\right)$ fm, and widths equal
to their errors, $\left(9,\, 3\right)$ fm. All the other EFT
parameters are assigned flat priors over ranges that correspond to, or
exceed, natural values: $0.001 \leq C^2_{1,2} \leq
1\,\mathrm{fm}^{-1}$, $0\leq r_{1,2} \leq 10\, \mathrm{fm} $
\cite{Phillips:1996ae, Wigner:1955zz}, $-1\leq \epsilon_1\leq 1$,
$-10\leq L_{1,2} \leq 10\, \mathrm{fm}$. We do, though, restrict the
parameter space by the requirement that there is no s-wave resonance
in $\bes$-p scattering below $0.6$ MeV.

To determine ${\rm pr}\left(\vec{g},\{\xi_i\} \vert D;T;I \right)$, we use a Markov chain Monte Carlo algorithm \cite{SiviaBayesian96} with  Metropolis-Hastings sampling \cite{Metropolis:1953am}, generating $2\times 10^4$ uncorrelated samples in  the $14$-dimensional (14d) $\vec{g}$ $\bigoplus$ $\{\xi_i\}$ space. Making histograms, e.g., over two parameters $g_1$ and $g_2$, produces the marginalized distribution, in that case: ${\rm pr} \left(g_{1}, g_{2} \vert D;T;I \right)=$  $\int {\rm pr} \left(\vec{g},\{\xi_i\} \vert D;T;I \right)\,$ $d\xi_1 \ldots d\xi_5 dg_3 \ldots dg_9$. Similarly, to compute the PDF of a quantity $F(\vec{g})$, e.g., $S(E; \vec{g})$, we construct ${\rm pr}\left(\bar{F}\vert D; T; I\right)$ $\equiv$ $\int {\rm pr} \left(\vec{g},\{\xi_i\} \vert D;T;I \right)$ $\delta(\bar{F}-F(\vec{g})) d\xi_1 \ldots d \xi_5 d\vec{g}$, and histogramming again suffices. 

\begin{figure}
\centering
\includegraphics[scale=1]{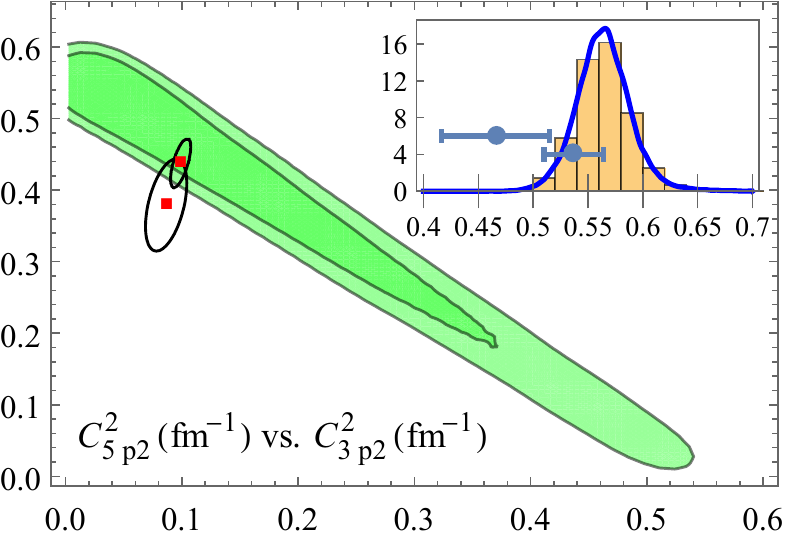}
\caption{(Color online.) 2d distribution for $C_1^2$ (x-axis) and
  $C_2^2$ (y-axis). Shading represents the 68\% and 95\% regions. The
  small circle and ellipse are the $1\sigma$ contours of an {\it ab
    initio} calculation \cite{Nollett:2011qf} and empirical results
  \cite{Tabacaru:2005hv}, with their best values marked as red
  squares. The inset is the histogram and the corresponding smoothed
  1d PDF of the quantity $[C_1^2+C_2^2]\times \mathrm{fm}$; the larger
  and smaller error bars show the empirical and {\it ab initio}
  values.} \label{fig:results1}
\end{figure}

{\em Constraints on parameters and the S-factor---} The tightest
parameter constraint we find is on the sum
$C_1^2+C_2^2=0.564(23)~\mathrm{fm}^{-1}$, which sets the overall scale
of $S(E)$~\footnote{The second moments of the MCMC sample distribution imply
that $C_1^2+ 0.94 C_2^2$ is best constrained, but we consider $C_1^2+
C_2^2$ for simplicity.}.  Fig.~\ref{fig:results1} shows contours of 68\% and 95\%
probability for the 2d joint PDF of the ANCs.  Neither ANC is
strongly constrained by itself, but they are strongly anticorrelated;
the 1d PDF of $C_1^2+ C_2^2$ is shown in the inset.  
The ellipses in
Fig.~\ref{fig:results1} show ANCs from an \textit{ab initio}
variational Monte Carlo calculation (the smaller ellipse)
\cite{Nollett:2011qf} \footnote{We recomputed the sampling errors of Ref.~\cite{Nollett:2011qf} in the basis of good $s$, taking more careful account of correlations between ANCs.} and inferred from transfer reactions by Tabacaru
\textit{et al.}  (larger ellipse) \cite{Tabacaru:2005hv}.  These are
also shown as error bars in the inset.  The \textit{ab initio} ANCs
shown compare well with the present results.  (The \textit{ab initio}
ANCs of Ref.~\cite{Navratil:2011sa} sum to $0.509~\mathrm{fm}^{-1}$ and
appear to be in moderate conflict.)   Tabacaru \textit{et al.}~recognized
that their result was $1\sigma$ to
$2\sigma$ below existing analyses of $S$-factor data; a
$1.8\sigma$ conflict remains in our analysis.  
We suggest that for
$^8$B the combination of simpler reaction mechanism, fewer
assumptions, and more precise cross sections makes the capture
reaction a better probe of ANCs than transfer reactions.

\begin{figure}

\centering

\includegraphics[scale=1]{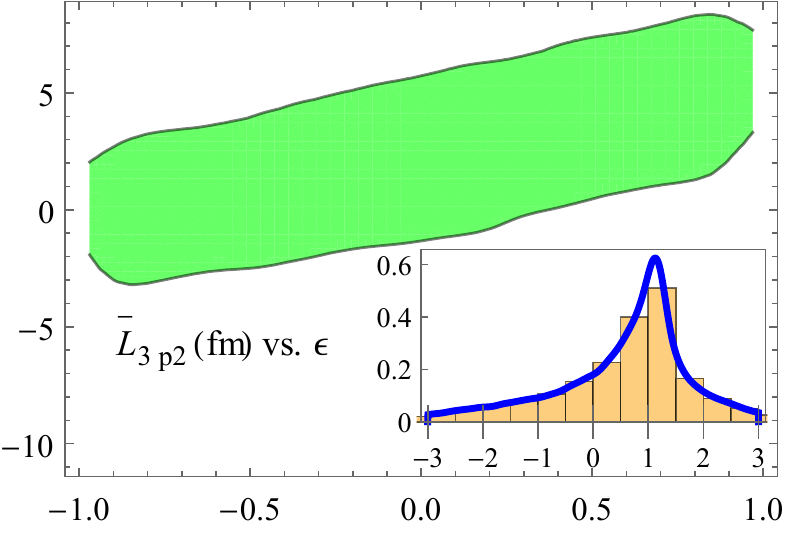}

\caption{(Color online.) 2d distribution for $\epsilon_1$ (x-axis) and $\bar{L}_1$ (y-axis). The shaded area is the 68\% region. The inset is the histogram and  corresponding smoothed 1d PDF of the quantity $0.33\, \bar{L}_1/\mathrm{fm} - \epsilon_1$.} \label{fig:results2}

\end{figure}

Fig.~\ref{fig:results2} depicts the 2d distribution of $\bar{L}_1$ and
$\epsilon_1$. There is a positive correlation: in $S(E)$ below the $\bes$-p inelastic threshold, the effect
of core excitation, here parameterized by $\epsilon_1$, can be traded
against the short-distance contribution to the spin-1 $E1$ matrix
element.  
The inset shows the $1$d distribution of the
quantity $0.33\, \bar{L}_1/\mathrm{fm} - \epsilon_1$, for which there
is a slight signal of a non-zero value. In contrast, the data 
prefers a positive $\bar{L}_2$: its 1d pdf yields a 68\% interval $-0.58~{\rm fm} < \bar{L}_2 < 7.94~{\rm fm}$. 

\begin{figure}

\centering

\includegraphics[scale=1]{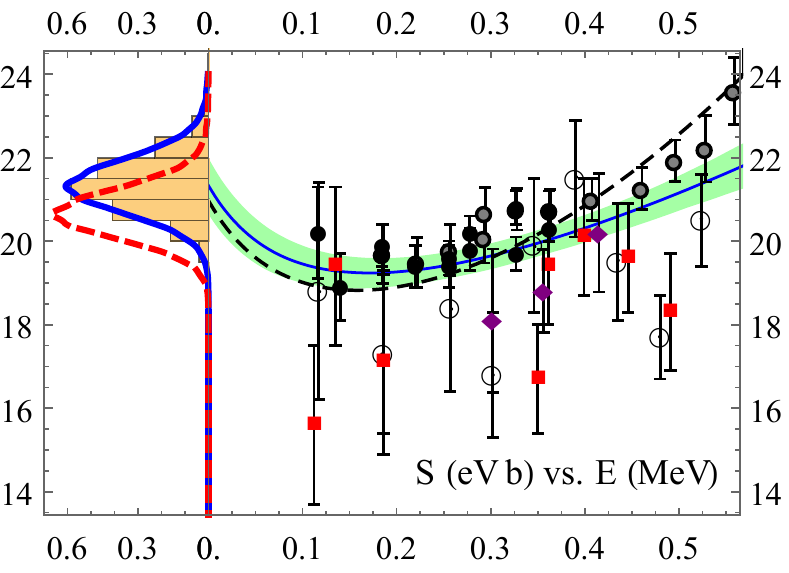}

\caption{(Color online.) The right panel shows the NLO $S$-factor at
  different energies, including the median values (solid blue
  curve). Shading indicates the 68\% interval. The dashed line is the
  LO result.  The data used for parameter determination are shown, but
  have not been rescaled in accord with our fitted $\{\xi_i\}$. They are:
Junghans {\it et.al.}, BE1 and BE3 \cite{Junghans:2010zz} (filled
black circle and filled grey circle), Filippone {\it et.al.,}
\cite{Filippone:1984us} (open circle), Baby {\it et.al.,}
\cite{Baby:2002hj, Baby:2002ju} (filled purple diamond), and Hammache
     {\it et.al.,} \cite{Hammache:1997rz, Hammache:2001tg} (filled red
     box). The left panel shows 1d PDFs for $S(0)$ (blue line and
     histogram) and $S(20~\mathrm{keV})$ (red-dashed
     line). } \label{fig:results3}

\end{figure}

We now compute the PDF of $S$ at many energies, and extract each median value (the thin solid blue line in Fig.~\ref{fig:results3}), and 68\% interval (shaded region in Fig.~\ref{fig:results3}). The PDFs for $S$ at $E=0$ and $20~\mathrm{keV}$ are singled out and shown on the left of the figure: the blue line and  histogram are for $E=0$ and the red-dashed line is for $E=20$ keV. We found choices of the EFT-parameter vector $\vec{g}$ (given in the supplemental material) that correspond to 
natural coefficients, produce curves close to the median $S(E)$
curve of Fig.~\ref{fig:results3}, and have large values of the posterior probability.

\begin{table}

\begin{ruledtabular} 

   \begin{tabular}{cccc}

	 	&$S$ (eV b) & $S'/S$ ($\mathrm{MeV}^{-1}$) & $S''/S$ ($\mathrm{MeV}^{-2} $) \\ \hline

 Median &21.33 [20.67]& $-1.82$ [$-1.34$] & 31.96 [22.30] \\

 $+\sigma$ & 0.66 [0.60] & 0.12 [0.12] & 0.33 [0.34]  \\

 $-\sigma$  & 0.69 [0.63] & 0.12 [0.12]  & 0.37 [0.38]\\  

\end{tabular} \caption{The median values of $S$, $S'/S$, and $S''/S$ at $E=0$ keV [$E=20$ keV], as well as the upper and lower limits of the (asymmetric) 68\% interval. The sampling errors are $0.02\%$, $0.07\%$, $0.01\%$ for median values, as estimated from $\left<X^2-\left<X\right>^2\right>^{1/2}/\sqrt{N}$ with $N=2 \times 10^4$.} \label{tab:SdSddS}

	\end{ruledtabular}

 \end{table}

{\it $S(20~keV)$ and the thermal reaction rate---}Table~\ref{tab:SdSddS} compiles median values and 68\% intervals for
the $S$-factor and its first two derivatives, $S^\prime/S$ and
$S^{\prime\prime}/S$, at $E=0$ and $20$ keV.
Ref.~\cite{Adelberger:2010qa} recommends $S(0)=20.8\pm 1.6$
\mbox{eV~b} (quadrature sum of theory and experimental
uncertainties). Our $S(0)$ is consistent with this, but the
uncertainty is more than a factor of two smaller.
Ref.~\cite{Adelberger:2010qa} also provides effective values of
$S^\prime/S=-1.5\pm 0.1~\mathrm{MeV}^{-1}$ and $S^{\prime
  \prime}/S=11\pm 4~\mathrm{MeV}^{-2}$.  These are not literal
derivatives but results of quadratic fits to several plausible models
over $0 < E < 50~\mathrm{keV}$, useful for applications.  Our values
are consistent, considering the large higher derivatives (rapidly
changing $S^{\prime\prime}$) left out of quadratic fits.

The important quantity for astrophysics is in fact not $S(E)$ but the
thermal reaction rate;
derivatives of $S(E)$ are used mainly in a customary
approximation to the rate
integral~\cite{caughlan62,rolfs-rodney,Adelberger:2010qa}.  By using
our $S^\prime$ and $S^{\prime\prime}$ in a Taylor series for $S(E)$
about $20$ keV, then regrouping terms and applying the approximation
formula, we find a rate (given numerically in the supplemental
material) that differs from numerical integration of our median $S(E)$
by only 0.01\% at temperature $T_9 \equiv T /( 10^9~\mathrm{K}) =
0.016$ (characteristic of the Sun), and 1\% at $T_9 = 0.1$ (relevant
for novae).

{\it How accurate is NLO?---}Our improved precision for $S(0)$ is achieved because, by appropriate choices of its nine parameters, NLO Halo EFT can represent all the models whose disagreement constitutes the 1.4 \mbox{eV b} uncertainty
quoted in Ref.~\cite{Adelberger:2010qa}---including the microscopic calculation of Ref.~\cite{Descouvemont:2004hh}.
Halo EFT matches their $S(E)$ and phase-shift curves with a precision of 1\% or better for $E < 0.5$ MeV, and thus spans
the space of models of $E1$ capture in the LER~\cite{Zhang:2015}. 

The LO curve shown in Fig.~\ref{fig:results3} employs values of $C_1$, $C_2$, $a_1$, and $a_2$ from the NLO fit. It differs from the NLO curve by $<2$\% at $E=0$, and by $< 10$\% at $E=0.5$ MeV. This rapid convergence suggests that the naive estimate of N2LO effects in the amplitude, $(k/\Lambda)^2\approx 4 \%$, is conservative. And indeed, we added a term with this $k$-dependence to the model, allowing a natural coefficient that was then marginalized over, and found that it shifted the median and error bars from the NLO result by at most $0.2\%$ in the LER. 
Finally, we estimate that direct $E2$ and $M1$ contributions to $S$ in the LER are less than $0.01\%$, and radiative corrections are around $0.01\%$.

{\em Summary---}
We used Halo EFT at next-to-leading order to determine
precisely the $^7\mathrm{Be}(p,\gamma)^8\mathrm{B}$ $S$-factor at
solar energies.  Halo EFT connects all low-energy models by a family of continuous parameters, and 
marginalization over those parameters represents marginalization over
all reasonable models of low-energy physics. 
Many of the individual EFT parameters are poorly determined by existing $S$-factor data,
at $E > 0.1$ MeV, but these data constrain their combinations
sufficiently that the extrapolated $S(20~\mathrm{keV})$ is determined
to 3\%.  We estimate that the impact of neglected higher-order terms in the EFT
on $S(0)$ is an order of magnitude smaller than this. 

Extension of the EFT to higher order and inclusion of couplings between 
s- and d-wave scattering states is not expected to reduce the uncertainty,
although it would provide slightly greater generality in matching
possible reaction mechanisms. There is, however, no indication in the
literature that coupling to $d$-waves is important for
$S(E)$~\cite{Descouvemont:2004hh} in the LER. Our analysis could perhaps be extended 
 to higher energies, but for $E > 0.5$ MeV,
accurate representation of $M1$ resonances is at least as important as
reliable calculations of the $E1$ transition.

 The most significant source of uncertainty in our extrapolant is, in fact, the $1$ keV uncertainty in the $\be$ proton-separation energy, which can shift  $S(20~\mathrm{keV})$ by approximately $0.75$\%. This could be eliminated by better mass measurements. Further significant improvement in $S(20~\mathrm{keV})$ for $\bes(p,\gamma)\be$ requires stronger constraints on EFT parameters.  Better determinations of $s$-wave
scattering parameters seem to be of limited utility. 
The ANCs affect the very-low-energy $S$-factor the most, and so more information on them, from
either \textit{ab initio} theory or capture/transfer data, would be useful. 

A number of other
radiative capture processes whose physics parallels $\bes(p,\gamma)\be$
are important in astrophysics. The formalism developed herein should be
applicable to many of them. 

{\em Acknowledgments---}
We thank Carl Brune for several useful discussions, and Barry Davids, Pierre Descouvemont, and Stefan Typel for sharing details of their calculations with us. 
We are grateful to the Institute for Nuclear Theory for support under Program INT-14-1, ``Universality in few-body systems: theoretical challenges and new directions", and Workshop INT-15-58W, ``Reactions and structure of exotic nuclei". During both we made significant progress on this project. 
X.Z.~and D.R.P.~acknowledge support from the US Department of Energy under grant DE-FG02-93ER-40756. K.M.N.~acknowledges support from the Institute of Nuclear and Particle Physics at Ohio University, and from U.S.~Department of Energy Awards No.~DE-SC 0010 300 and No.~DE-FG02-09ER41621 at the University of South Carolina.

\bibliographystyle{apsrev}
\bibliography{nuclear_reaction}

\widetext

\section{Supplemental material}

\subsection{Common-mode errors for experimental data}

The quoted common-mode errors for Junghans {\it et al.}, sets BE1 and BE3,
\cite{Junghans:2010zz}, Filippone {\it et.al.,}
\cite{Filippone:1984us}, Baby {\it et.al.,} \cite{Baby:2002hj,
  Baby:2002ju}, and the Hammache {\it et.al.,} 1998 data
set~\cite{Hammache:1997rz} are $2.7\%$, $2.3\%$, $11.25\%$, $5\%$, and
$2.2\%$, respectively. The data of Ref.~\cite{Hammache:2001tg} are a
measurement of the absolute $S(186~\mathrm{keV})$ and of the ratios
$S(135~\mathrm{keV})/S(186~\mathrm{keV})$ and
$S(112~\mathrm{keV})/S(186~\mathrm{keV})$. We treat each of these
three quantities as one data point, so they do not need a CME.

\subsection{EFT details}

The modified-effective-range expansion for s-wave $\bes$-p scattering is:
\begin{equation}
p\left(\cot \delta_s(E) - i\right)\, \frac{2\pi \eta}{e^{2\pi\eta}-1} =-\frac{1}{a_{s}} + \frac{1}{2} r_{s} p^{2} -2\kc  H(\eta).
 \end{equation} 
Here  $H(\eta) =\psi(i\eta)+{1}/{(2i\eta)}-\ln(i\eta)$, $\eta \equiv k_C/p$, $k_C \equiv Z_{\bes} Z_p \alpha_{em} m_R$ with $m_R$ the $\bes$-p reduced mass, $p=\sqrt{2 m_R E}$, and $\psi$ the digamma function~\cite{MathHandBook1}.
 
An EFT parameter set that gives a good fit---as mentioned in the main text---is listed in Table~\ref{tab:aEFTfit}.  

\begin{table}
 \begin{ruledtabular} 
   \begin{tabular}{ccccccccc}
	 $C_{(\P{3}{2})}^{2}$ ($\mathrm{fm}^{-1}$)  & $a_{(\S{3}{1})}$ (fm) & $r_{(\S{3}{1})}$ (fm) & $\epsilon_{1} $ & $ \overline{L}_{1}$ (fm) & $C_{(\P{5}{2})}^{2}$ ($\mathrm{fm}^{-1}$)  & $a_{(\S{5}{2})}$ (fm) & $r_{(\S{5}{2})}$ (fm) & $ \overline{L}_{2}$ (fm) \\ \hline
0.2336 & 24.44 & 3.774 & -0.04022 & 1.641 & 0.3269 & -7.680 & 3.713 & 0.1612 \\ 
      \end{tabular}      
\caption{ A representative EFT parameter set that gives a curve almost
  on the top of the median value curve (solid blue) in
  Fig.~\ref{fig:results3}. The LO curve (dashed black) uses the LO
  parameters listed here, with the strictly NLO parameters set to zero.
  Because the parameter space is very degenerate, many such parameter
  sets could be given that have similar $S(E)$ curves but very
  different parameter values.} \label{tab:aEFTfit}
		\end{ruledtabular}
\end{table}
 
\subsection{Results for $S$-factor and thermal reaction rate}

\begin{table}
 \begin{ruledtabular} 
   \begin{tabular}{cccc}
 	$E$ (MeV) &  Median (eV b)  & $-\sigma$ (eV b) & $+\sigma$ (eV b) \\ \hline
	0. & 21.33 & 0.69 & 0.66 \\
 0.01 & 20.97 & 0.65 & 0.63 \\
 0.02 & 20.67 & 0.63 & 0.60 \\
 0.03 & 20.42 & 0.60 & 0.58 \\
 0.04 & 20.20 & 0.57 & 0.55 \\
 0.05 & 20.02 & 0.55 & 0.53 \\
 0.1 & 19.46 & 0.45 & 0.44 \\
 0.2 & 19.27 & 0.34 & 0.34 \\
 0.3 & 19.65 & 0.32 & 0.30 \\
 0.4 & 20.32 & 0.35 & 0.31 \\
 0.5 & 21.16 & 0.42 & 0.41 \\
\end{tabular}      \caption{The median values and 68\% interval bounds for $S$ in the energy range from 0 to 0.5 MeV. At each energy point, the histogram of $S$ is drawn from the Monte-Carlo simulated ensemble and then is used to compute the median and the bounds. } \label{tab:s0to0.5MeV}
		\end{ruledtabular} 
\end{table}

The median values and 68\% interval bounds for $S$ in 10 keV intervals to 50 keV and then in 100 keV intervals to 500 keV is listed in Table~\ref{tab:s0to0.5MeV}. 

Regrouping the Taylor series for 
$S(E)$ about $20$ keV into a quadratic and applying the approximations of Refs.~\cite{caughlan62,rolfs-rodney} yields
\begin{equation}
N_A \langle \sigma v \rangle=\frac{2.7648 \times10^5}{T_9^{2/3}} \exp\left(\frac{-10.26256}{T_9^{1/3}}\right)
\times (1 + 0.0406 T_9^{1/3}  - 0.5099 T_9^{2/3} - 0.1449 T_9
   +0.9397 T_9^{4/3} + 0.6791 T_9^{5/3}),
\label{eq:thermalreactionrate}
\end{equation}
in units of 
$\mathrm{cm^3\,s^{-1}\,mol^{-1}}$, where $N_A$ is Avogadro's number.
Up to $T_9=0.6$, the lower and upper
limits of the 
68\% interval for $S(E)$ produce a numerically integrated rate that is
 $0.969 (1+0.0576 T_9-0.0593 T_9^2)$ and $1.030
(1-0.05 T_9 +0.0511 T_9^2)$ times that of Eq.~(\ref{eq:thermalreactionrate}). 
At $T_9 \gtrsim 0.7$
energies beyond the LER, and hence resonances, come into play and so these results no longer hold. We know of no
astrophysical environment with such high $T_9$ where
$^7\mathrm{Be}(p,\gamma)^8\mathrm{B}$ matters. 
\narrowtext

\end{document}